\documentclass[namedreferences,hyperref,optionalrh]{spr-sola}
\usepackage{graphicx}        
\usepackage{color}           

\chardef\us=`\_

\begin{document}

\begin{frontmatter}
\title{Models of Collisionless Quasineutral Solar Wind Current Sheets}


\author[addressref={aff1},
corref,
email={sb388@st-andrews.ac.uk}]{\inits{S.}\fnm{Sophie}~\snm{Boswell}\orcid{0009-0004-1531-3760}}
\author[addressref=aff1,
email={tn3@st-andrews.ac.uk}]{\inits{T.}\fnm{Thomas}~\snm{Neukirch}\orcid{0000-0002-7597-4980}}
\author[addressref=aff2]{\inits{A. N.}\fnm{Anton}~\snm{Artemyev}\orcid{0000-0002-9404-4105}}
\author[addressref=aff3]{\inits{I. Y.}\fnm{Ivan}~\snm{Vasko}\orcid{0000-0002-4974-4786}}
\author[addressref={aff4,aff5,aff6}]{\inits{O.}\fnm{Oliver}~\snm{Allanson}\orcid{0000-0003-2353-8586}}
\address[id=aff1]{School of Mathematics and Statistics, University of St. Andrews, St. Andrews, UK}
\address[id=aff2]{Department of Earth, Planetary, and Space Sciences, University of California, Los Angeles, Los Angeles, California, USA}
\address[id=aff3]{William B. Hanson Center for Space Sciences, University of Texas at Dallas, Richardson, TX, USA}
\address[id=aff4]{Space Environment and Radio Engineering, School of Engineering, University of Birmingham, Birmingham, UK}
\address[id=aff5]{Department of Earth and Environmental Sciences, University of Exeter, Penryn, United Kingdom}
\address[id=aff6]{Department of Mathematics and Statistics, University of Exeter, Exeter, United Kingdom}

\runningauthor{Boswell et al.}
\runningtitle{Quasineutral Solar Wind Current Sheets}

\begin{abstract}
In situ measurements of kinetic scale current sheets in the solar wind show that they are often approximately force-free although the plasma 
beta
is of order one. They frequently display systematic asymmetric and anti-correlated 
spatial variations of their particle density and temperature across the current sheet, leaving the plasma pressure essentially uniform. These observations of asymmetries have 
previously been modelled theoretically by adding additional terms to both the ion and electron distribution functions 
of self-consistent force-free collisionless current sheet models with constant density and temperature profiles. In this paper we present the results of a modification of these models in which only the electron distribution function has a term added to it, whereas the ion distribution function is kept as a thermal (Maxwellian) distribution function. In this case the nonlinear quasineutrality condition no longer has a simple analytical solution and therefore has to be solved alongside Amp\`{e}re's law. We find that while the magnetic field remains approximately force-free, the non-zero quasineutral electric field gives rise to an additional spatial substructure of the plasma density inside the current sheet. We briefly discuss the potential relation between our theoretical findings and current sheet observations.


\end{abstract}
\end{frontmatter}

\section{Introduction}
     \label{S-Introduction} 




Current sheets are ubiquitous features of space and
astrophysical plasmas. They play a fundamental role in many activity processes, e.g. magnetic reconnection.
The present study is motivated by kinetic scale current sheets observed in the solar wind throughout 
the inner heliosphere \citep[e.g.][]{vasko:etal2021,vasko:etal2022,lotekar:etal2022}.

We particularly focus on one-dimensional models of currents sheets 
for which the magnetic field component
perpendicular to the current sheet vanishes ("tangential discontinuities" as opposed to "rotational 
discontinuities" for which the perpendicular magnetic field component is non-zero). We note that
even if the perpendicular magnetic field component is non-zero, it is 
often so small
compared to the magnetic 
field magnitude within the current sheet
that it cannot be determined reliably even using multi-spacecraft observations \citep[see, e.g.][]{Knetter:etal2004:JGR,Wang:etal2024:JGR}.
In situ measurements show that these current sheets are often 
approximately one-dimensional and force-free,
i.e. that the current density is aligned with the magnetic field. This implies that
the magnitude of the magnetic field remains approximately constant while the magnetic field
direction rotates across the current sheet \citep[e.g][]{Burlaga77,Lepping&Behannon86,Neugebauer06,Paschmann13:angeo}.

Statistical analysis of a large number of current sheet observations
\citep[][]{artemyev2019kinetic,Artemyev-2019b}
recently revealed systematic anticorrelated spatial variations of the plasma density $n$, and the
ion and electron temperatures $T_{i,e}$ across the current sheet, leaving the plasma pressure
uniform.
\citet{neukirch2020kinetic} presented self-consistent collisionless equilibrium models for
current sheets with the observed properties. These models were based on a modification
of the equilibrium distribution functions
for the force-free Harris sheet \citep[][]{harrison2009one, Neukirch-2009}. 
Whereas in the usual force-free case the ion distribution function can be taken to 
simply be a thermal
(Maxwellian)
distribution, for mathematical convenience \citet{neukirch2020kinetic}
had to modify both the electron and the ion distribution functions. In this paper we investigate
the consequences of leaving the ion distribution function unchanged from the force-free case
and modifying only the electron distribution function.

The paper is structured as follows. In Section \ref{S-Background Theory} we will
give a brief overview of the general theory of the collisionless force-free
Harris sheet and the modifications made to the distribution functions by 
\citet{neukirch2020kinetic}. 
In Section \ref{S-Quasineutral Theory} we present how
the mathematical problem  of finding current sheet solutions changes if only the 
electron distribution function is modified, in particular the resulting nonlinear
quasineutrality condition linking the electric potential and the magnetic vector
potential. We then show how one can find solutions by different methods, including 
solving the full nonlinear problem with numerical methods.
In Section 
\ref{S-Discussion and Conclusions}, we present a summary and our conclusions.

\section{Background Theory}
     \label{S-Background Theory} 

We briefly summarise the background theory of 
one-dimensional
collisionless force-free current sheets in this section
\citep[for a more detailed account see e.g.][]{Neukirch-2017}.
We use Cartesian coordinates $x$, $y$, $z$ and make the choice that spatial variations 
occur in the $z$-direction only. We investigate current sheets that only have non-vanishing 
magnetic field components
in the $x$- and $y$-directions, i.e. $\mathbf{B}(z)= [B_x(z), B_y(z), 0]$. 
Solutions to the stationary Vlasov equation, i.e. equilibrium distribution functions,
only depend on constants of motion related to the charged particle dynamics in the electromagnetic fields 
\citep[see e.g.][Chapter 6]{Schindler-book06}. For the chosen set-up the 
three known constants of motion are associated with the symmetries of the problem, i.e.
time-invariance, 
translational-invariance, and are 
\begin{itemize}
\item the Hamiltonian (total energy) for the particle species, $s$, given by $H_{s}=m_{s}v^{2}/2 + q_{s}\Phi$, and 
\item the 
$x$- and $y$-components of the canonical momentum,
$p_{x,s} = m_{s}v_{x} + q_{s}A_{x}$ and $p_{y,s}=m_{s}v_{y} + q_{s}A_{y}$.
\end{itemize}
Here, $v_{x}$, $v_{y}$, and $v_{z}$ are the $x$-, $y$-, and $z$-components of the velocity, and $v^2=v_{x}^2+v_{y}^2+v_{z}^2$. Also, $\Phi(z)$ is the electric potential, $A_x(z)$ and $A_y(z)$ are the $x$- and $y$-components of
the vector potential, and $m_s$ and $q_s$ are the mass and electric charge of particle species $s$, respectively.
In this paper we assume that the plasma consists only of electrons ($s=e$) and protons ($s=i$).

Self-consistent solutions to the Vlasov-Maxwell equations are found by calculating the charge and current densities
from the equilibrium distribution functions $F_s(H_s, p_{x,s}, p_{y,s})$, and
using them as source terms in the inhomogeneous Maxwell equations. The homogeneous Maxwell
equations 
(Faraday's law and the solenoidal condition for the magnetic field) are 
automatically satisfied by the fact that the electromagnetic fields are time-independent and the use of the electric and vector potentials.
For quasineutral plasmas, Gauss's law for the 
electric 
potential is well approximated by the quasineutral
condition
\begin{equation}
    \sigma(\Phi,A_x,A_y) = e[n_i(\Phi,A_x,A_y) - n_e(\Phi,A_x,A_y)] =0 ,
    \label{eq:qn-general}
\end{equation}
where $\sigma$ is the charge density, $e$ is the elementary charge, and $n_i$ and $n_e$ are the proton
and electron particle densities defined by the zeroth order velocity moments of the distribution functions,
\begin{equation}
    n_s = \int F_s(H_s, p_{x,s}, p_{y,s})\, \mathrm{d}^3 v.
    \label{eq:zero-v-moment}
\end{equation}
This is coupled with Amp\`ere's law for the vector potential,
\begin{eqnarray}
    -\frac{\mathrm{d}^2 A_x}{\mathrm{d} z^2} &=& \mu_0 j_x(\Phi,A_x,A_y), \label{eq:ampere-x}\\
   - \frac{\mathrm{d}^2 A_y}{\mathrm{d} z^2} &=&  \mu_0 j_y(\Phi,A_x,A_y) ,\label{eq:ampere-y}
\end{eqnarray}
where $\mu_{0}$ is the permeability of free space and the current density components are defined by
\begin{equation}
    j_k (\Phi,A_x,A_y) = \sum_s q_s \int v_k \,F_s(H_s, p_{x,s}, p_{y,s})\, \mathrm{d}^3 v, \quad k=x, y.
\end{equation}

The 
model 
we use as a starting point for our investigations
in this paper is the force-free Harris sheet for which a self-consistent 
distribution function was first found by \citet[][]{harrison2009one}. The magnetic field
is given by
\begin{equation}
\mathbf{B}(z)=B_{0}
\left[\tanh(z/L),\frac{1}{\cosh(z/L)},0\right]
\label{magfield},
\end{equation}
such that $B_{x}^{2} + B_{y}^{2} = B_{0}^{2} = const$. 
Here, $B_0$ is the magnitude of the magnetic field and $L$ is the typical current sheet width.
%
%
 
Electron and ion distribution functions self-consistent with this magnetic field 
profile are given by \citep[see e.g.][]{harrison2009one,neukirch2020kinetic}
%
%
\begin{eqnarray}
  F_{e}(H_{e},p_{x,e},p_{y,e}) &=& \frac{n_{0}}{\sqrt{2\pi v_{th,e}^{2}}^{3}}
    \exp(-\beta_{e}H_{e}) \times  \nonumber  \\
 &&  \;   
 \left[ -\frac{1}{2}\cos(\beta_{e}u_{0}p_{x,e}) 
       \exp\left(\frac{u_{0}^{2}}{2 v_{th,e}^{2}}\right)
 +
\right.                            \nonumber  \\
 && \qquad \;
              \left. 
              \exp(\beta_{e}u_{0}p_{y,e})
              \exp\left(-\frac{u_{0}^{2}}{2 v_{th,e}
              ^{2}}\right)
              +b \right],\label{DFe}\\
    F_{i}(H_{i},p_{x,i},p_{y,i})&=& \frac{n_{0,i}}{\sqrt{2\pi v_{th,i}^{2}}^{3}}\exp(-\beta_{i}H_{i}),\label{DFi}
\end{eqnarray}
where 
$n_{0}$ and $n_{0,i}$ are typical particle densities, $u_{0}$ is a constant velocity parameter, 
$b$ is a dimensionless parameter with
$b+\frac{1}{2}$ being 
the
value of the plasma beta,
and the usual notation of $\beta_{s}=(k_{B}T_{s})^{-1}$ for the inverse temperature has been used, with $k_B$ the Boltzmann
constant. Also, $v_{th,s}^2 = (m_s\beta_s)^{-1}$ is the usual thermal velocity of particle species $s$.

For completeness and later reference we also note that for this current sheet equilibrium one finds that
the electric potential $\Phi(z)$ is identically zero, and the vector potential components 
$A_{x}(z)$ and $A_{y}(z)$ 
are given by 
\begin{eqnarray}
A_{x,ff}(z) &= &B_{0}L\arctan[\sinh(z/L)], \\
A_{y,ff}(z) &= & -B_{0}L\ln[\cosh(z/L)]. 
\end{eqnarray}
We remark that the only gauge transformation allowed for the one-dimensional current sheets we investigate 
in this paper compatible with the symmetry properties we use are additive constants to the electric and vector potentials. We fix the gauge by assuming that $A_{x,ff}(z)$ is an odd function of $z$ and $A_{y,ff}(z)$ is an even function of $z$ with $A_{y,ff}(0)=0$.
More details regarding this current sheet model and the modification that will be
described next can be found in Appendix \ref{S-appendix-CS}.

To be able model the observed spatial asymmetries in the
electron and ion
density 
and temperature \citep{artemyev2019kinetic,Artemyev-2019b}
\citet[][]{neukirch2020kinetic} added the following additional term to both the ion and electron distribution functions
\begin{equation}
    \Delta F_{s}= \delta n_{s} \left( \frac{\kappa_{s}}{2\pi 
    v_{th,s}^{2}} \right)^{\frac{3}{2}}\beta_{s}u_{0}p_{x,s}\left(\frac{5}{2}-\kappa_{s}\beta_{s}H_{s}\right)
    e^{-\kappa_{s}\beta_{s}H_{s}},\label{DFaddterm}
\end{equation}
where $\delta n_{s}$  is a constant with the dimension of a particle density, and $\kappa_{s}$ is a dimensionless parameter.

The magnetic field remains as given in Equation \ref{magfield}, i.e. force-free, and so the force-free condition detailed in \cite{harrison2009some} that requires the relevant component of the pressure tensor, in this case the $zz-$ component, $
{P}_{zz}$, to be constant across the sheet is upheld
(see Appendix \ref{S-appendix-CS} for details).
However, the additional term in the distribution functions leads to an asymmetric 
contribution 
to the 
particle density of the 
form \citep[][]{neukirch2020kinetic}
\begin{equation}
\Delta 
n_{s} = \epsilon n_{0}\frac{2A_{x,ff}}{B_{0}L}
\label{eq:Deltans}
\end{equation}
where 
$\delta n_{e} = -
\delta n_i (\beta_i/\beta_e) =\epsilon n_{0}$.
In combination with a constant pressure, the non-uniform plasma density corresponds to an anti-correlated temperature 
asymmetry
across the sheet. 

One interesting point to note is that the ion distribution function 
for the force-free Harris sheet with constant density and
temperature can be chosen to simply be a thermal distribution function (Maxwellian). 
However, \citet[][]{neukirch2020kinetic} had to modify the ion distribution function in the same way as the 
electron distribution function in order to have $\Phi =0 $ as an exact solution of the 
quasineutrality
condition,
and to leave the magnetic field structure unchanged
(see Appendix \ref{S-appendix-CS}). In this paper we will investigate
how the solution changes if one assumes that the ion distribution function is kept as a purely thermal distribution.

\section{Quasineutral Theory}
     \label{S-Quasineutral Theory} 

\subsection{Basic Equations}     

When using the thermal ion distribution function given by Equation \ref{DFi}
together with the full electron distribution function given by the sum of Equation \ref{DFe}
and the additional population given by Equation \ref{DFaddterm},
getting a self-consistent solution becomes more
complicated and
generally requires the use of numerical methods, as we will explain below. The  quasineutrality condition
%
takes the form
\begin{equation}
    n_{0,i}e^{-e\beta_{i}\Phi} - n_{0}
    \left[ e^{e\beta_{e}\Phi}
    N(A_x,A_y)
      -\epsilon e\beta_{e}u_{0}A_{x}(1 + \kappa_{e}e\beta_{e}\Phi)e^{\kappa_{e}e\beta_{e}\Phi} \right] = 0,
      \label{QN}   
\end{equation}
where
\begin{equation}
   N(A_x,A_y) = b -
    \frac{1}{2}\cos
    (e\beta_{e}u_{0}A_{x})+ 
    e^{-e\beta_{e}u_{0}A_{y}}.
    \label{eq-defN}
\end{equation}
In Equation \ref{QN}, we have used $q_i = e$ and $q_e=-e$.

The quasineutrality condition 
is coupled with Amp\`ere's law
\begin{eqnarray}
        -\frac{d^{2}A_{x}}{dz^{2}} &=& \mu_{0}en_{0}u_{0}e^{e\beta_{e}\Phi}\frac{1}{2}\sin(e\beta_{e}u_{0}A_{x}) - \epsilon\mu_{0}en_{0}u_{0}e\beta_{e}\Phi e^{\kappa_{e}e\beta_{e}\Phi},\label{Ampx}\\
         -\frac{d^{2}A_{y}}{dz^{2}} &=& -\mu_{0}en_{0}u_{0}e^{e\beta_{e}\Phi}e^{-e\beta_{e}u_{0}A_{y}},\label{Ampy}
\end{eqnarray}
defining a differential-algebraic (DAE) system of equations for $\Phi$, $A_x$, and $A_y$. 
%
For $\epsilon = 0$, we obtain the force-free case where $\Phi =0$, $A_{x}=A_{x,ff}$, and $A_{y}=A_{y,ff}$ (Appendix \ref{S-appendix-CS}).
The various parameters satisfy the 
relations \citep[see e.g.][]{Neukirch-2017,neukirch2020kinetic}
\begin{eqnarray}
     N(A_{x,ff},A_{y,ff})&=& b + \frac{1}{2},\label{rel1}\\
    n_{0,i} &=& n_{0}\left(b+\frac{1}{2}\right),\label{rel2}\\
    \frac{B_{0}L}{2} &=& -\frac{1}{e\beta_{e}u_{0}},\label{rel3}\\
    \frac{B_{0}}{L} &=& -\mu_{0}en_{0}u_{0},\label{rel4}\\
    \frac{B_{0}^{2}}{2\mu_{0}} &=& \frac{n_{0}}{\beta_{e}},\label{rel5}\\
    L^{2} &=& \frac{2}{\mu_{0}e^{2}\beta_{e}n_{0}u_{0}^{2}}.\label{rel6}
\end{eqnarray}
Equation \ref{QN}
is not solved by $\Phi =0$ and the force-free vector potential components $A_{x,ff}$ 
and $A_{y,ff}$. 
This 
can, for example, be
seen by considering 
the
final term, which depends linearly on $A_{x}$ 
and therefore would still 
vary
with $z$
when $\Phi =0$. 
Therefore, to get self-consistent solutions for the
electric potential and the magnetic field, the set of
equations \ref{QN} - \ref{Ampy} have to be solved
either numerically or using
analytical approximations.

\subsection{Expanding the Electric Potential for 
Small $\epsilon$}
    \label{S-Expansion Method}

As noted in \citet{neukirch2020kinetic}, the value of the
parameter $\epsilon$ needed for the theoretical density asymmetry to match the observed level of asymmetry is quite small ($\epsilon \approx 0.05$). 
This suggests that a
method to obtain sufficiently accurate approximate solutions
could be based on an expansion in $\epsilon$ of the problem defined by 
Equations \ref{QN} - \ref{Ampy}.

\citet[][]{neukirch2022kinetic} 
expanded the
electric potential 
about $\Phi = 0$, such that $\Phi = \epsilon \Phi_{1} + ...$ to  obtain an approximate solution
of Equation \ref{QN} up to order $\epsilon^{1}$. 
Under the assumption
that the magnetic field, and hence magnetic vector potential, remain unchanged, one obtains the following relation between the electric potential and the (force-free) vector potential:
\begin{equation}
    e\beta_{e}\Phi_{1} = -\frac{2}{\left(b + \frac{1}{2}\right)\left(1 + \frac{\beta_{i}}{\beta_{e}}\right)}\frac{A_{x,ff}}{B_{0}L}.\label{Phicrude}
\end{equation}
When substituted into the corresponding expression for the asymmetric
electron density term $\Delta n_{e}$ this results in
\begin{equation}
    \Delta n_{e} = 
    \epsilon n_{0}\frac{\frac{\beta{i}}{\beta_e}}{\left(1 +\frac{\beta{i}}{\beta_e}\right)}\frac{2A_{x,ff}}{B_{0}L}\label{necrude}.
\end{equation}
Comparing Equation \ref{necrude} with Equation \ref{eq:Deltans} one notices 
that 
they differ by the
factor 
$\frac{\frac{\beta{i}}{\beta_e}}{1 +\frac{\beta{i}}{\beta_e}}$ in the approximate expression 
for the 
quasineutral
case.
For $\beta_{i}=\beta_{e}$, for example, this implies
that for this level of approximation 
the asymmetric density component 
would differ from that in Equation \ref{eq:Deltans} by a factor $1/2$.

However, from Amp\`{e}re's law (Equations \ref{Ampx} and \ref{Ampy})
it is clear that the assumption that the magnetic field remains unchanged at order $\epsilon^{1}$ is inconsistent,
because the current density depends on the electric potential $\Phi$. Hence, 
it will be modified for $\Phi \ne 0$ and
the vector potential for the force-free magnetic field will no longer be an
exact solution of Amp\`{e}re's law. One therefore
has to improve this ``crude" approximation
by extending the 
expansion to include the
vector potential components, which will be
done in the next subsections. 

The reason 
we include the ``crude" approximation presented in this
subsection is 
that the resulting particle densities and temperatures
have the same spatial structure
as those found 
by \citet{neukirch2020kinetic},
albeit scaled
by a constant factor. This
allows us to illustrate the difference between
the spatial variation of the original
model and the more consistent solutions that
will be presented in the next two subsections. The
``crude" approximation results are therefore included
in Figures \ref{F-linear} and \ref{F-nonlinear}
for comparison.

\subsection{
Full Expansion Method for Small $\epsilon$: Linear Case}
    \label{S-Linear Theory}


We now generalise the expansion method of \citet[][]{neukirch2022kinetic} to include the 
magnetic vector potential components $A_{x}$ and $A_{y}$.
In 
normalised form, with $\bar{z} = \frac{z}{L}$, $\bar{\Phi}=e\beta_{e}\Phi$, and all vector potential components normalised by $B_{0}L$, Equations \ref{QN} - \ref{Ampy} become
%
\begin{eqnarray}
    0 &=& \left(b+\frac{1}{2}\right)e^{-\frac{\beta_{i}}{\beta_{e}}\bar{\Phi}}-e^{\bar{\Phi}}N(\bar{A}_{x},\bar{A}_{y}) -2\epsilon(1+\kappa_{e}\bar{\Phi})e^{\kappa_{e}\bar{\Phi}}\bar{A}_{x},\label{normQN}\\
    %
    %
    N(\bar{A}_{x},\bar{A}_{y}) &=&b-\frac{1}{2}\cos(2\bar{A}_{x})+e^{2\bar{A}_{y}},\label{normN}\\
    \frac{d^{2}\bar{A}_{x}}{d\bar{z}^{2}} &=& 
    -\frac{1}{2}e^{\bar{\Phi}}\sin(2\bar{A}_{x})
    -\epsilon \bar{\Phi}e^{\kappa_{e}\bar{\Phi}},\label{normAmpx}\\
    \frac{d^{2}\bar{A}_{y}}{d\bar{z}^{2}} &=& -e^{\bar{\Phi}}e^{2\bar{A}_{y}}.\label{normAmpy}
\end{eqnarray}
As in Section \ref{S-Expansion Method}, 
the electric potential is expanded about 
$\bar{\Phi} = 0$.
The magnetic vector potential components are expanded about their 
force-free solutions 
$\bar{A}_{x,ff}$ and $\bar{A}_{y,ff}$,
such that 
$\bar{A_{x}} = \bar{A}_{x,ff} + \epsilon \bar{A}_{x1} +...$ and 
$\bar{A}_{y} = \bar{A}_{y,ff} + \epsilon \bar{A}_{y1} +...$.

The order $1 (= \epsilon^0)$ terms then correspond to the force-free case and
at
order $\epsilon^{1}$ we obtain
\begin{eqnarray}
    \bar{\Phi}_{1} &=& -\frac{2}{\left(b+\frac{1}{2}\right)\left(\frac{\beta_{i}}{\beta_{e}}+1\right)}\left[ 
    \bar{A}_{x,ff} + \frac{\sinh(\bar{z})}{\cosh^{2}(\bar{z})}\bar{A}_{x1} + \frac{1}{\cosh^{2}(\bar{z})}\bar{A}_{y1}\right],\label{linearPhi}\\
    \frac{d^{2}\bar{A}_{x1}}{d\bar{z}^{2}} &=& \left(1 -\frac{2}{\cosh^{2}(\bar{z})}\right)\bar{A}_{x1} 
    - \frac{\sinh(\bar{z})}{\cosh^{2}(\bar{z})}\bar{\Phi}_{1},\label{linearAmpx}\\
    \frac{d^{2}\bar{A}_{y1}}{d\bar{z}^{2}} &=& -\frac{1}{\cosh^{2}(\bar{z})}[2\bar{A}_{y1} + \bar{\Phi}_{1}].\label{linearAmpy}    
\end{eqnarray}
Using Equation \ref{linearPhi} on the right-hand sides of Equations \ref{linearAmpx} and \ref{linearAmpy} results
in two coupled linear inhomogeneous 
second-order differential equations for 
$\bar{A}_{x1}$ and $\bar{A}_{y1}$.
We emphasize that solving Eqs.\ \ref{linearAmpx} and \ref{linearAmpy} will result in electromagnetic potentials, and thus magnetic fields, that are self-consistent up to order $\epsilon$.

This system of equations has to be completed by 
initial and/or boundary conditions. The fact that the current density has to 
have a finite extent in 
$\bar{z}$
results in the requirement that 
$\bar{A}_{x1}
\to 0$ as $|\bar{z}| \to
\infty$. The reason for this is the first term on the right hand side of the 
$\bar{A}_{x1}$
equation. Because
$\bar{A}_{x1}$ 
tends
to a constant (here equal to zero) as 
$|\bar{z}| \to \infty$
we also ensure that 
$\bar{B}_{y1} 
\to 0$ in that limit and therefore
$\bar{B}_{y} 
\to 0$ as well.
For 
$\bar{A}_{y1}$ 
we
impose the condition $\mathrm{d} 
\bar{A}_{y1}/d \bar{z} 
\to 0$ for $|\bar{z}| \to \infty$. This ensure that 
$\bar{B}_{x1} 
\to 0$ in this limit and that 
$\bar{B}_{x}$ 
approaches
its force-free value outside the current sheet.

\begin{figure}
\centerline{\hspace*{0.015\textwidth}
         \includegraphics[width=0.515\textwidth,clip=]{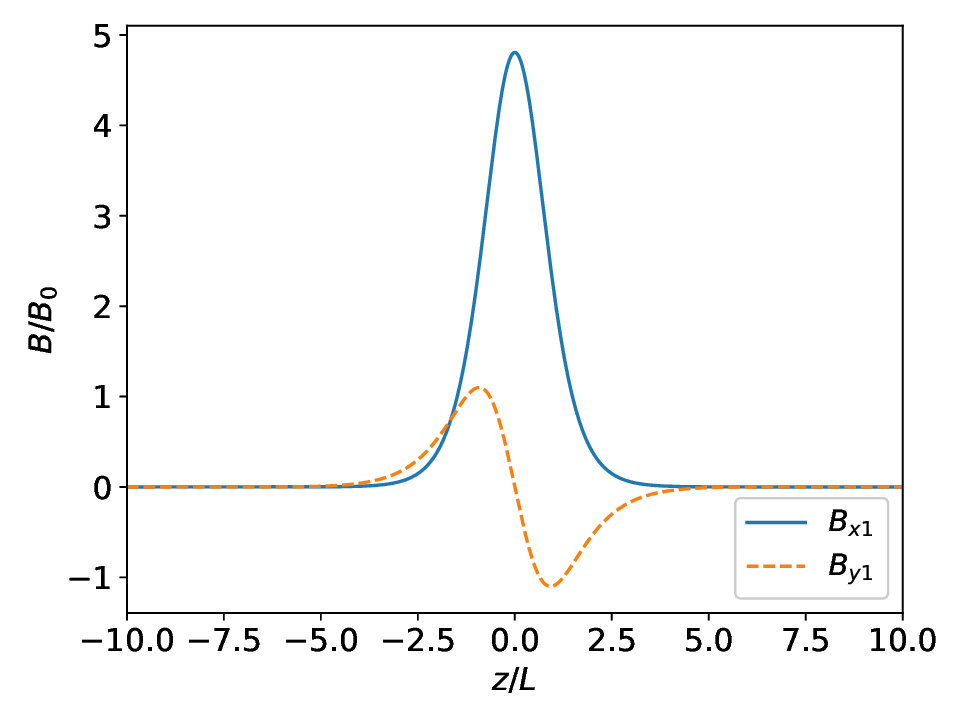}
         \hspace*{-0.03\textwidth}
         \includegraphics[width=0.515\textwidth,clip=]{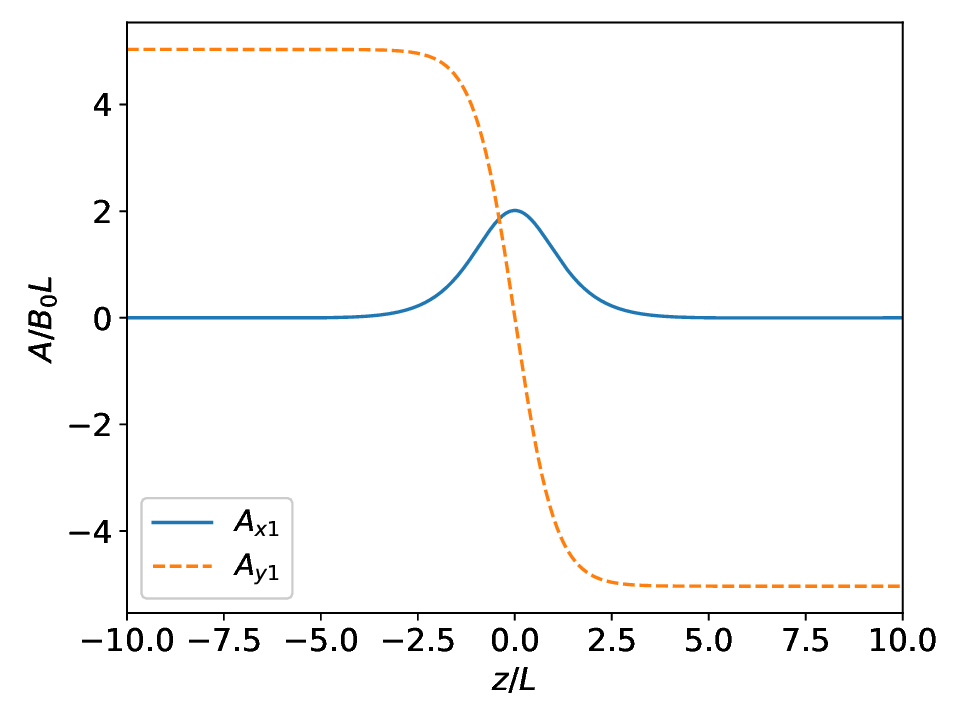}\hspace*{-0.03\textwidth}

        }
\vspace{-0.351\textwidth}   
\centerline{\Large \bf     
\hspace{0.1 \textwidth}  \color{black}{(a)}
\hspace{0.415\textwidth}  \color{black}{(b)}
   \hfill}
\vspace{0.31\textwidth}    
\centerline{\hspace*{0.015\textwidth}
         \includegraphics[width=0.515\textwidth,clip=]{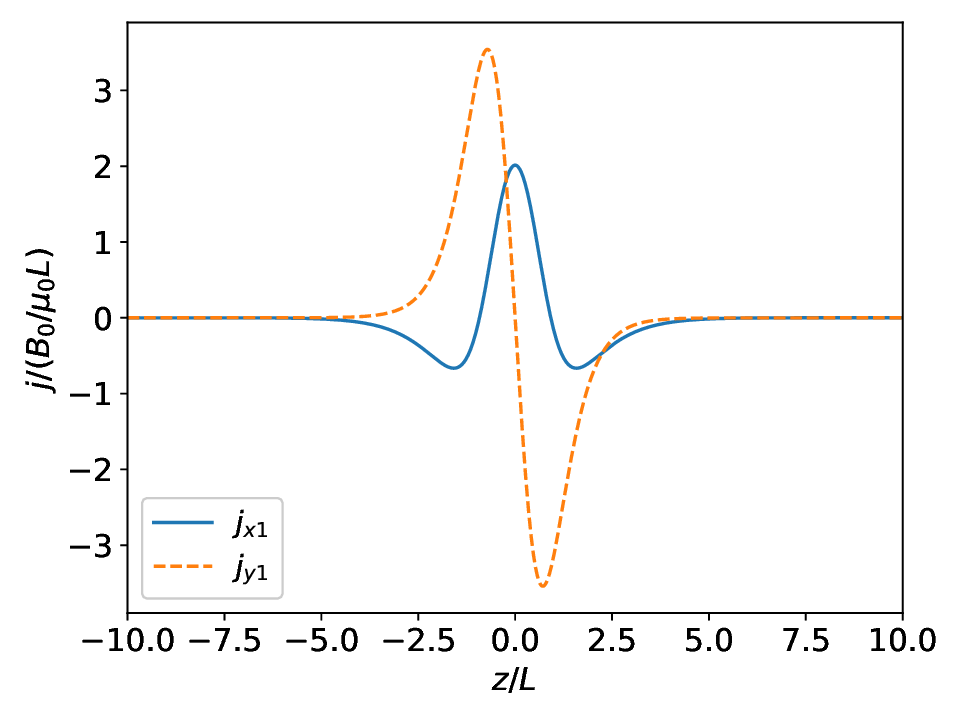}
         \hspace*{-0.03\textwidth}
         }
\vspace{-0.351\textwidth}   
\centerline{\Large \bf     
\hspace{0.35 \textwidth} \color{black}{(c)}
\hfill}
\vspace{0.31\textwidth}    
\caption{Solutions to the linear ODE system for the (a) magnetic field, (b) magnetic vector potential, and (c) current density components
for the parameter values
$b=0.9$,
$T_e/T_i=1.0$,
$\kappa_e=1.1$, and
$\epsilon = 0.05$. }
\label{F-linear_expansion}
\end{figure}

\begin{figure}
\centerline{\hspace*{0.015\textwidth}
         \includegraphics[width=0.515\textwidth,clip=]{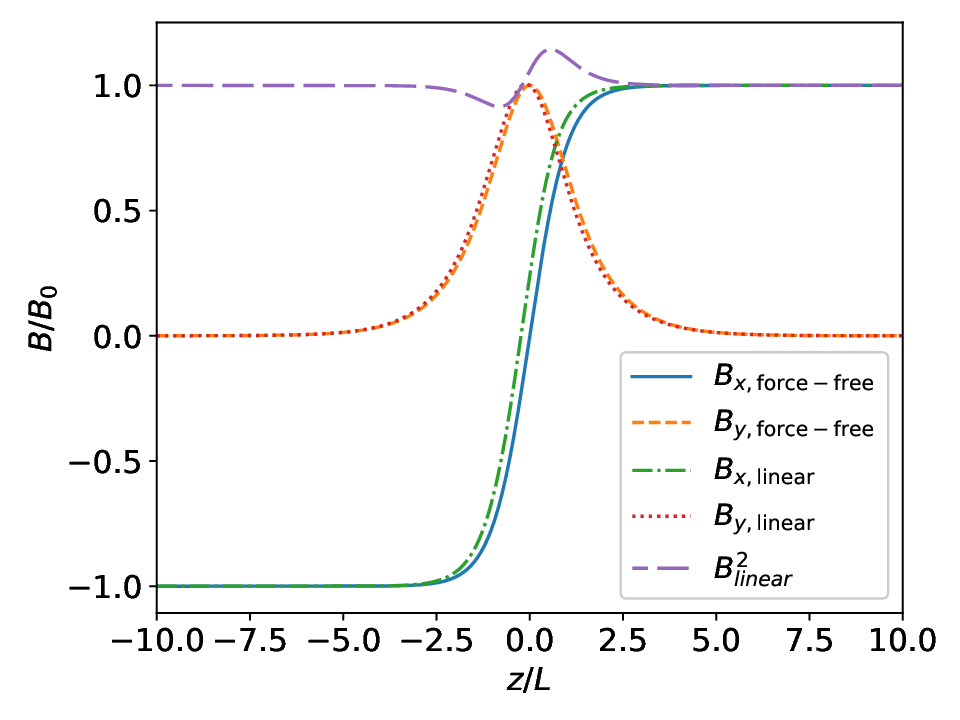}
         \hspace*{-0.03\textwidth}
         \includegraphics[width=0.515\textwidth,clip=]{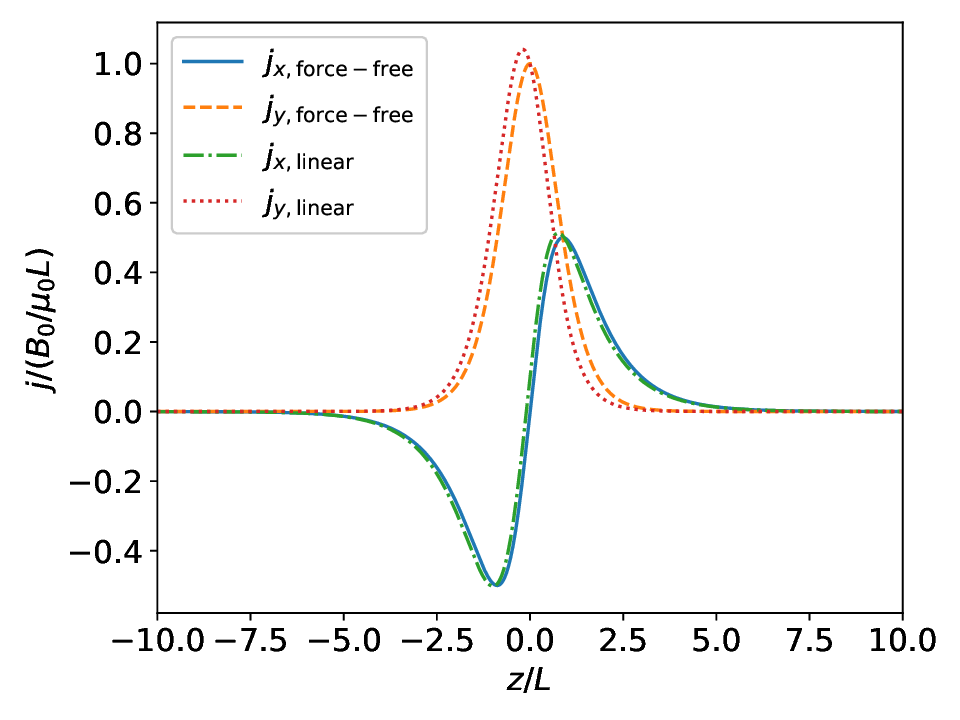}
        }
\vspace{-0.351\textwidth}   
\centerline{\Large \bf     
\hspace{0.375 \textwidth}  \color{black}{(a)}
\hspace{0.415\textwidth}  \color{black}{(b)}
   \hfill}
\vspace{0.31\textwidth}    
\centerline{\hspace*{0.015\textwidth}
         \includegraphics[width=0.515\textwidth,clip=]{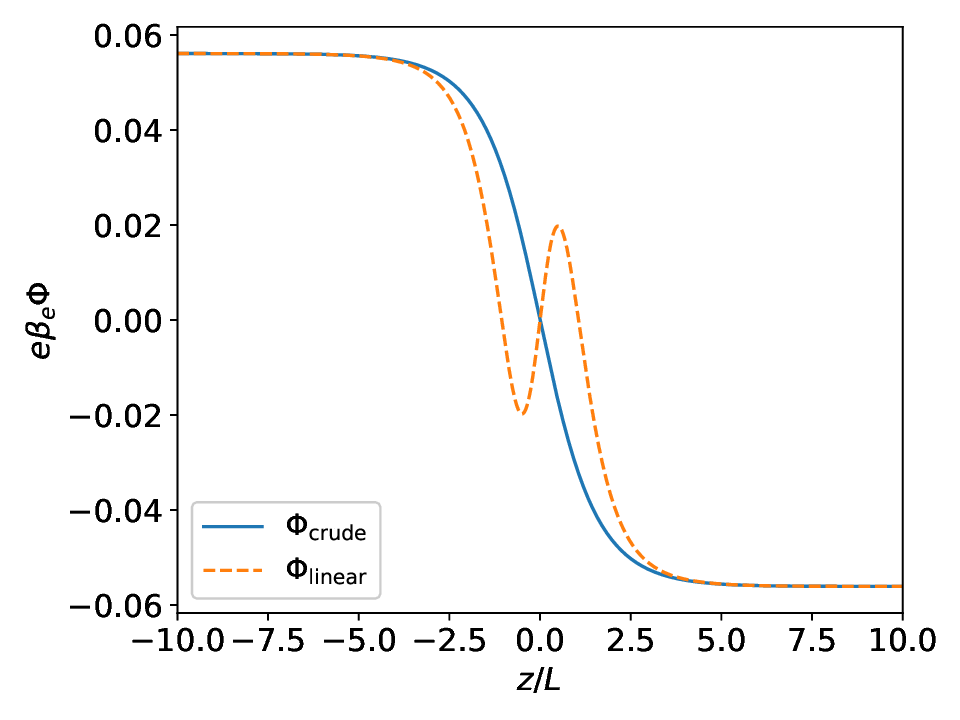}
         \hspace*{-0.03\textwidth}
         \includegraphics[width=0.515\textwidth,clip=]{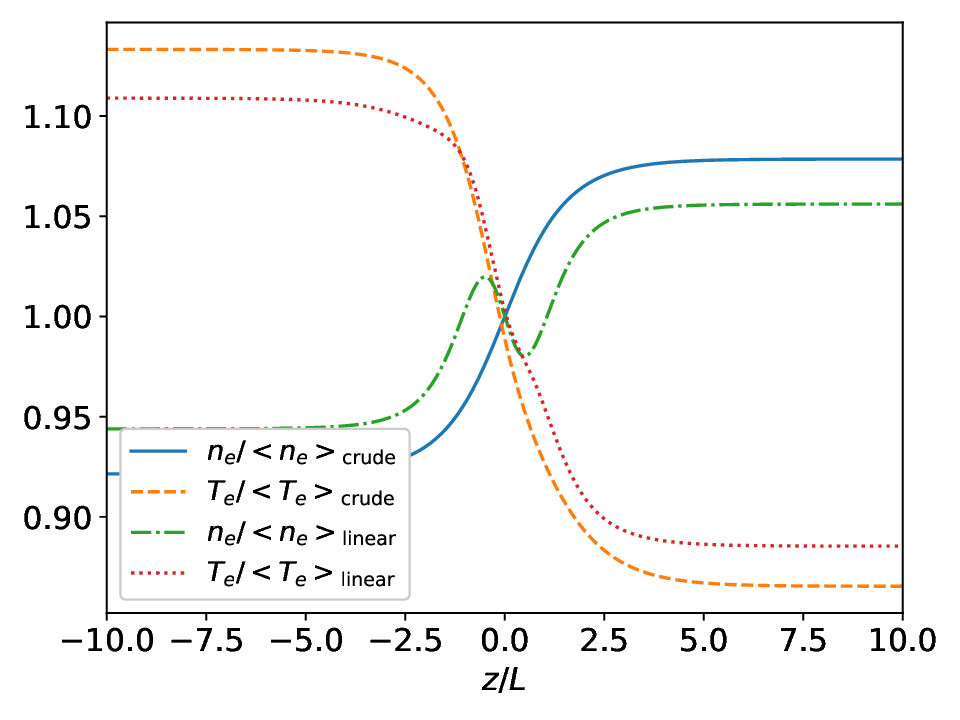}
        }
\vspace{-0.351\textwidth}   
\centerline{\Large \bf     
\hspace{0.375 \textwidth} \color{black}{(c)}
\hspace{0.415\textwidth}  \color{black}{(d)}
   \hfill}
\vspace{0.31\textwidth}    
              
\caption{Resulting linear solutions of the 
(a) magnetic field profile, 
(b) current density profile, 
(c) electric potential, and (d) density and temperature asymmetries
for the parameter values
$b=0.9$,
$T_e/T_i=1.0$,
$\kappa_e=1.1$, and
$\epsilon = 0.05$. 
Each panel compares the linear solution with the 
``crude" approximation for the same parameter values 
(see Section \ref{S-Expansion Method}).
The magnetic field and current density components for
the ``crude" approximation are those of the force-free solution.}
\label{F-linear}
\end{figure}

Equations \ref{linearPhi} - \ref{linearAmpy}
were solved
numerically for the same 
parameter values
used in
\citet[][]{neukirch2020kinetic},
i.e. $b=0.9$ (which corresponds to a plasma beta of $1.4$),
$T_{e}/T_{i}=1.0$, $\kappa_{e}=1.1$, and $\epsilon=0.05$. 
The resulting solutions to the linear ODE system for the magnetic field, magnetic vector potential, and current density components are shown in Figure \ref{F-linear_expansion}.
The resulting profiles for the magnetic field, current density, 
electric potential, and particle density and temperature 
are shown
in Figure 
\ref{F-linear}.
The 
two terms (magnetic pressure and $zz$-component
of the pressure tensor) contributing 
to
the force balance equation for the 
linear solution up to order $\epsilon$ are shown in Figure \ref{F-lin_force_balance}. As stated in Appendix 
\ref{S-appendix-CS} force balance is equivalent to
the total pressure being constant with respect to the $z$-coordinate, and Figure \ref{F-lin_force_balance} shows
that this is case here.
We remark that we have also 
calculated solutions for a number
of other parameter combinations, but found that
the solutions look very similar apart from very simple
differences due to, for example, scaling of amplitudes.

\begin{figure} 
\centerline{\includegraphics[width=0.7\textwidth,clip=]{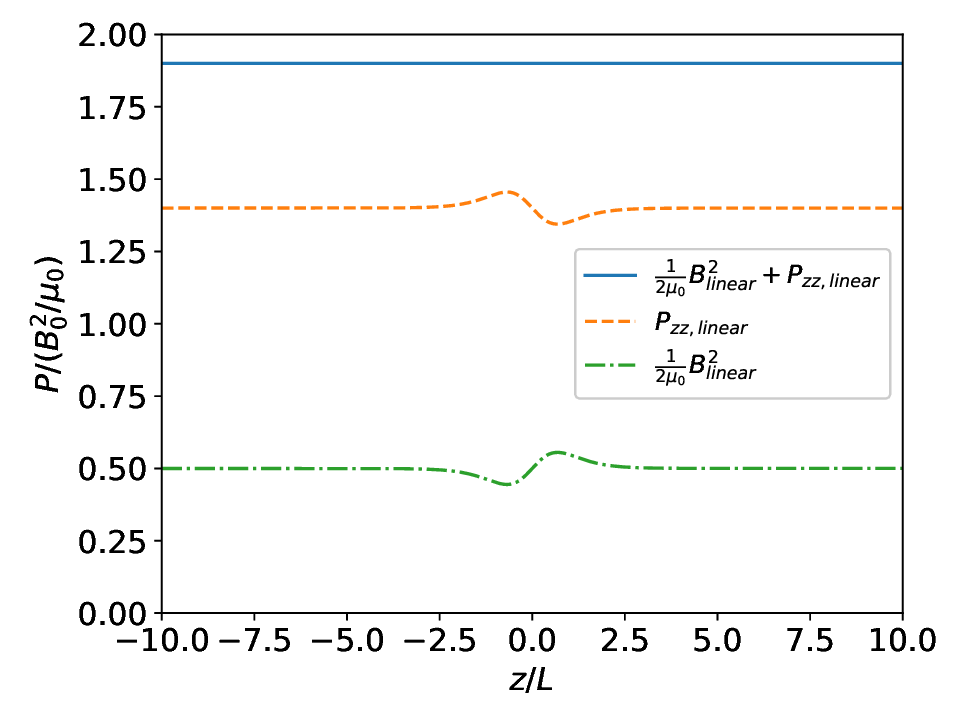}}
\vspace{-0.67\textwidth}   

\vspace{0.65\textwidth}

        \caption{
        Variation with $z$ of the magnetic pressure 
        and $zz$-component of the pressure tensor 
        for the 
        full 
        linear solution, i.e. including all terms up to 
        order $\epsilon$. Although both the magnetic pressure and the $zz$-component of the pressure tensor vary with $z$
        their sum is constant, which shows that force balance holds up to order $\epsilon$
        (for details on force balance see Appendix \ref{S-appendix-CS}).}
\label{F-lin_force_balance}
\end{figure}

The magnetic field and current density profiles are 
broadly consistent with 
those 
of the force-free case 
\citep[e.g.][]{harrison2009one,Neukirch-2009}
with a
small shift
of the new profiles in $z$ noticable. This is caused by 
the additional terms in the current density and the
need for the first order quantities to satisfy 
the imposed boundary and initial conditions.
As a consequence of this the magnetic field is no longer completely force-free, as 
can be seen by the deviation of 
$B^2(\bar{z})$ 
from a straight line within the current sheet.
The amplitude of this deviation is, however, small.

Somewhat surprisingly, 
the
electric potential and the electron density and temperature profiles
display additional spatial structures
around the centre of the sheet which are 
different from 
those found 
under the assumption that the magnetic field is unchanged
at first order in $\epsilon$.
Hence,
the
inclusion of $A_{x}$ and $A_{y}$ in the expansion leads to 
particle density profiles 
that are not simply asymmetrical, but also have some 
spatial substructure within the current sheet.
\subsection{Nonlinear Theory}
    \label{S-Non-Linear Theory}
To corroborate the results found 
by
using the expansion 
method, we also solved the full 
nonlinear
problem as formulated in Equations \ref{QN} - \ref{Ampy}.
Due to the
nonlinear
nature of this
differential algebraic system of equations (DAE) 
solutions must be found numerically. 

The normalised quasineutrality condition given by Equation 
\ref{normQN}
is a transcendental equation and cannot be 
solved
analytically
for $\bar{\Phi}$.
Hence, 
one needs to solve the full set of equations given by
Equations \ref{QN} - \ref{Ampy}.
To do so,
we 
follow the method
used by \citet{catapano:etal2015}. This method
replaces the algebraic 
quasineutrality
condition (Equation \ref{normQN})
by a differential equation for the
electric potential $\bar{\Phi}$.
This differential equation
is obtained by simply differentiating 
the 
quasineutrality
condition with respect to
$\bar{z}$, leading to the ODE
%
%
\begin{eqnarray}
    \frac{d \bar{\Phi}}{d\bar{z}} &=& \left(\left[e^{\bar{\Phi}}\sin(2\bar{A}_{x})+2\epsilon(1+\kappa_{e}\bar{\Phi})e^{\kappa_{e}\bar{\Phi}}\right]\frac{d \bar{A}_{x}}{d \bar{z}}  +\left[2e^{\bar{\Phi}}e^{2\bar{A}_{y}}\right]\frac{d \bar{A}_{y}}{d \bar{z}}\right) \nonumber \\ && \hspace{2em} \cdot\left(-\left[b+\frac{1}{2}\right]\frac{\beta_{i}}{\beta_{e}}e^{-\frac{\beta_{i}}{\beta_{e}}\bar{\Phi}} -e^{\bar{\Phi}}
    N(\bar{A}_{x},\bar{A}_{y})
    \right. \nonumber \\ && 
    \hspace{12em}\left. 
    -2\epsilon\kappa_{e}\bar{A}_{x}(2+\kappa_{e}\bar{\Phi})e^{\kappa_{e}\bar{\Phi}}\right)^{-1}\label{dPhidz}.
\end{eqnarray}
%
%
Equation \ref{dPhidz}
is solved 
alongside
Equations \ref{normAmpx} and 
\ref{normAmpy} to obtain a solution to the full 
nonlinear
problem. To facilitate the use of 
similar
boundary
conditions 
as
in the linear case, 
we define the differences 
$\Delta\bar{A}_x$ and $\Delta\bar{A}_y$ 
between the vector potential components and
their force-free counterparts by
$\bar{A}_x = \bar{A}_{x,ff} + \Delta\bar{A}_x$, 
$\bar{A}_y = \bar{A}_{y,ff} + \Delta\bar{A}_y$, 
and
reformulate the set of differential equations for this
difference (Appendix \ref{S-appendix}). 
The boundary conditions used for the nonlinear 
calculations are
$\mathrm{d}
\Delta\bar{A}_x
/\mathrm{d} \bar{z} \to 0$ and 
$\mathrm{d}
\Delta\bar{A}_y
/\mathrm{d} \bar{z} \to 0$ for
$\left|\bar{z}\right|
\to \infty$,
which are similar to those
of the linear case and ensure that one obtains a 
spatially confined
current density structure, i.e. a current sheet. 
Due to the additional 
differential equation for the electric potential one
has to impose a fifth condition, which has been chosen
as 
$\Delta\bar{A}_y 
\to 0 $ for $\bar{z} \to - \infty$.

\begin{figure}
\centerline{\hspace*{0.015\textwidth}
         \includegraphics[width=0.515\textwidth,clip=]{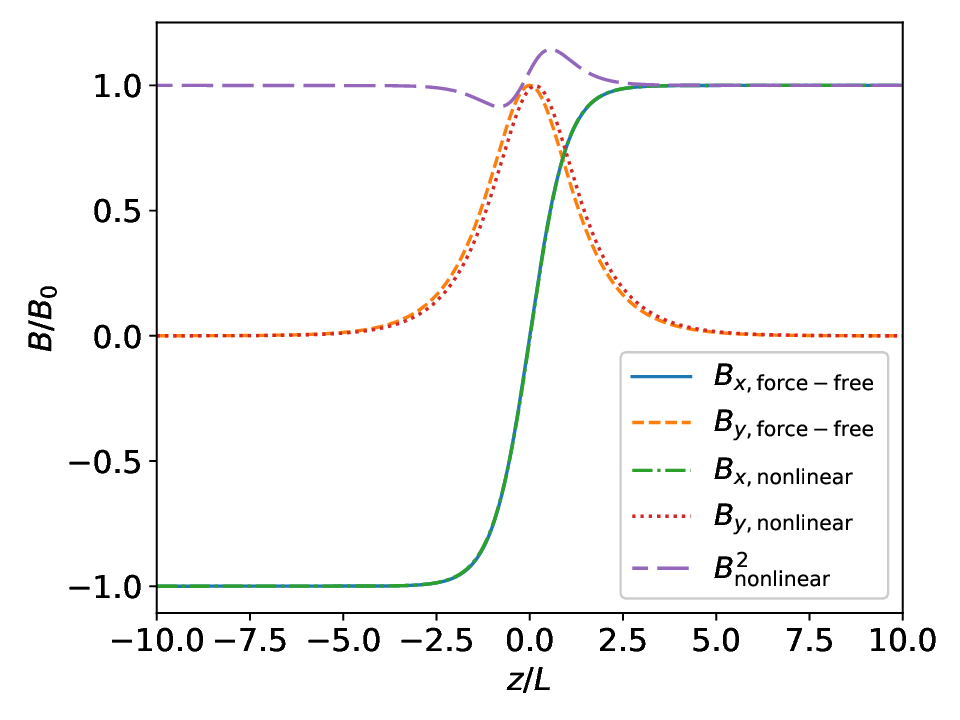}
         \hspace*{-0.03\textwidth}
         \includegraphics[width=0.515\textwidth,clip=]{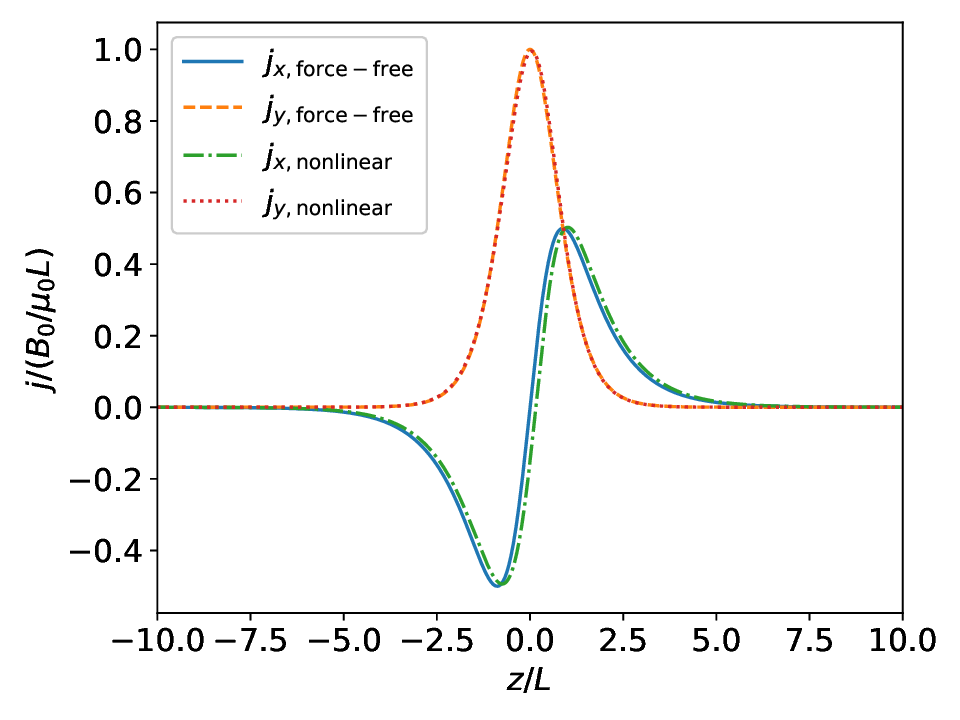}
        }
\vspace{-0.351\textwidth}   
\centerline{\Large \bf     
\hspace{0.375 \textwidth}  \color{black}{(a)}
\hspace{0.415\textwidth}  \color{black}{(b)}
   \hfill}
\vspace{0.31\textwidth}    
\centerline{\hspace*{0.015\textwidth}
         \includegraphics[width=0.515\textwidth,clip=]{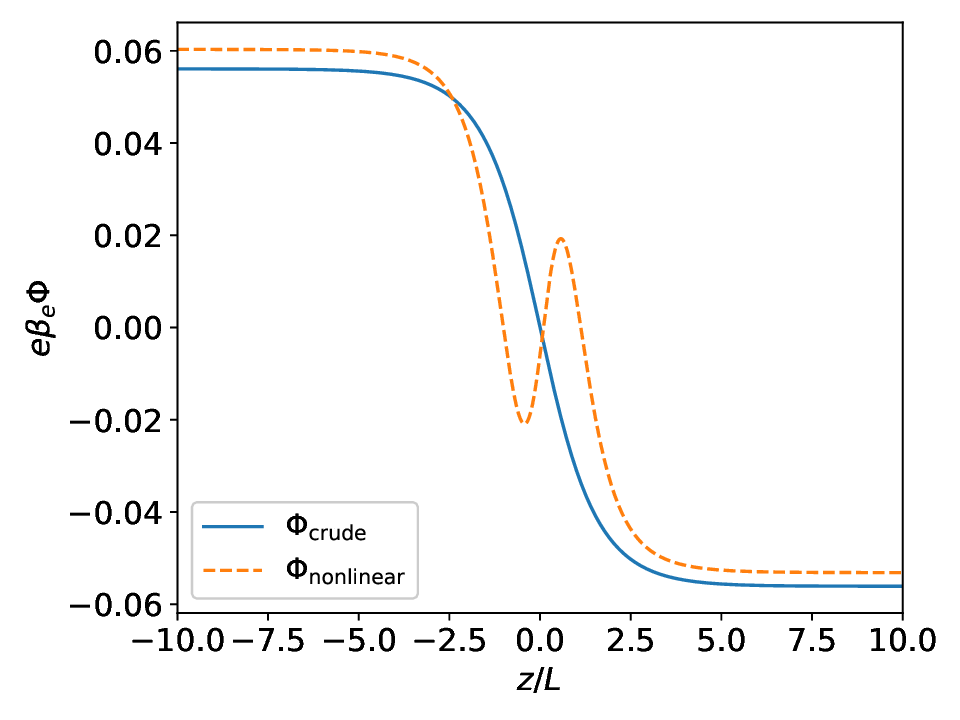}
         \hspace*{-0.03\textwidth}
         \includegraphics[width=0.515\textwidth,clip=]{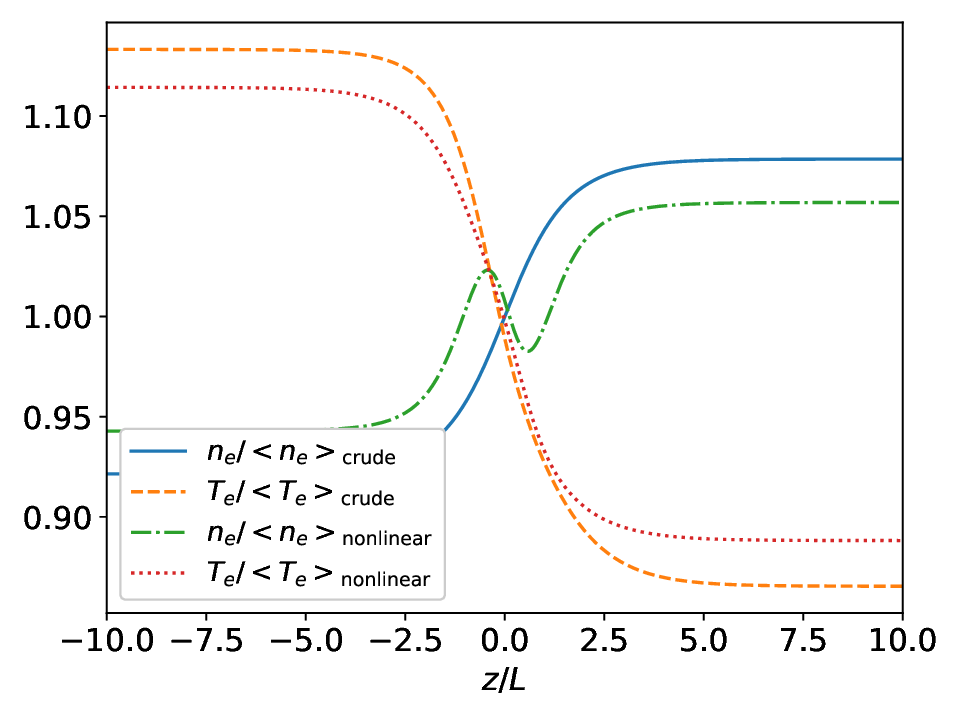}
        }
\vspace{-0.351\textwidth}   
\centerline{\Large \bf     
\hspace{0.375 \textwidth} \color{black}{(c)}
\hspace{0.415\textwidth}  \color{black}{(d)}
   \hfill}
\vspace{0.31\textwidth}    
              
\caption{Resulting 
nonlinear 
solutions of the (a) magnetic field profile, (b) current density profile, (c) electric potential, and (d) density and temperature asymmetries
for the parameter values 
$b=0.9$,
$T_e/T_i=1.0$,
$\kappa_e=1.1$, and
$\epsilon = 0.05$.
Each panel compares the 
nonlinear
solution with the ``crude" approximation 
(Section \ref{S-Expansion Method}).
The magnetic field and current density components
for the ``crude" approximation case are those of the force-free solution.}
\label{F-nonlinear}
\end{figure}

\begin{figure} 
\centerline{\includegraphics[width=0.7\textwidth,clip=]{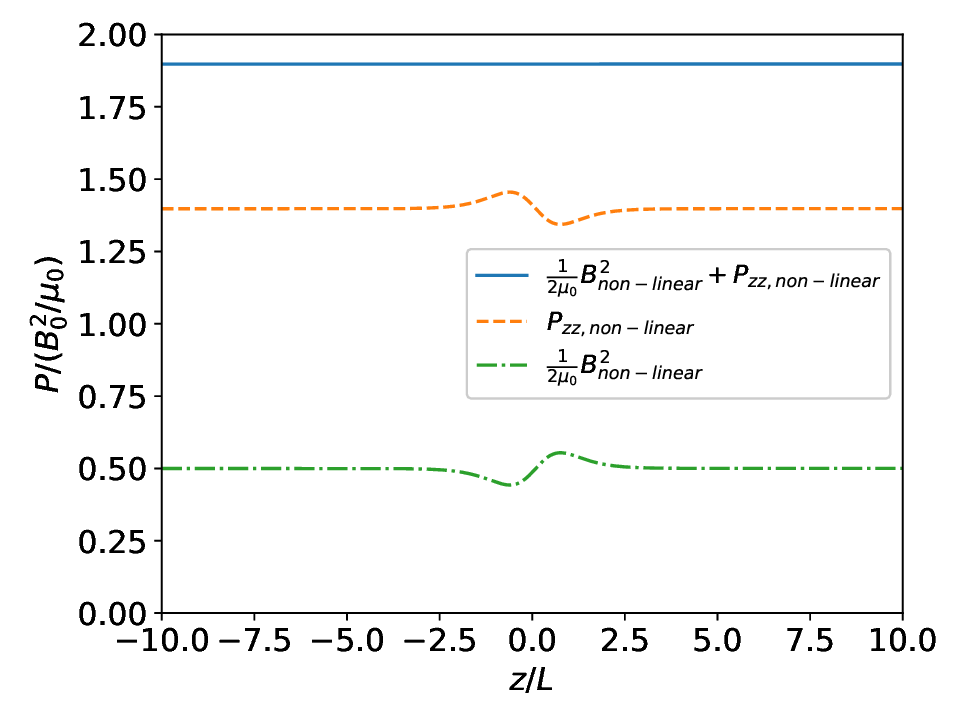}}
\vspace{-0.67\textwidth}   

\vspace{0.65\textwidth}

        \caption{
        Variation with $z$ of the magnetic pressure, $zz$-component of the pressure tensor and their sum (total pressure) for the full non-linear solution.
        Both the magnetic pressure and the $zz$-component of the pressure tensor vary, but 
        the total pressure is constant implying that 
        the current sheet is in force balance
        (for details on force balance see Appendix \ref{S-appendix-CS}).}
\label{F-nonlin_force_balance}
\end{figure}

We show in Figure \ref{F-nonlinear} the results 
of the numerical calculation for the self-consistent magnetic field, 
current density, 
electric potential, and 
particle density and temperature. In Figure \ref{F-nonlin_force_balance} we show 
the variation across the current sheet of the magnetic 
pressure, the $zz$-component of the pressure tensor and
their sum (total pressure)
for the full non-linear problem. 
As Figure \ref{F-nonlin_force_balance} shows
the total pressure is constant across the sheet which
implies that the current sheet is in force balance.

In 
these calculations
the same parameter values
as in 
Figures \ref{F-linear_expansion}-\ref{F-lin_force_balance}
(${b=0.9}$,
$T_{e}/T_{i}=1.0$, $\kappa_{e}=1.1$, $\epsilon=0.05$) have been
used. Solutions for a number
of other parameter combinations 
have also been calculated, but were found to
be structurally very similar, albeit with somewhat
different amplitudes, for example.


As in the linear case, the magnetic field and current density profiles 
do not deviate much
from
the force-free case 
\citep[][]{harrison2009one,Neukirch-2009}
although they again are slighty 
different, 
which is due to the additional dependence of
the current density on the electric potential in
combination with the imposed boundary conditions. Both
the magnetic field and current density variations
are largely identical for the linear and the nonlinear
cases, showing the consistency of the results for
both cases.

One also sees that the electric potential and electron density and temperature profiles show 
spatial variations
very similar to
those found in the linear case. 
The nonlinear calculations are thus fully consistent 
with the results of the linear case and 
corroborate these. The consistency between the
linear expansion method and the nonlinear case also
shows that the ``crude" approximation fails to 
provide the correct solution for the electric 
potential and the particle
density. In view of the relatively minor differences
between the force-free magnetic field and
the magnetic fields found in the linear expansion method
and the nonlinear case, this is a somewhat surprising
result, but shows that making consistent assumptions
is important.

\section{Summary and Conclusions}
    \label{S-Discussion and Conclusions}

In this paper we have presented a generalisation of
the 
theoretical approach by 
\citet{neukirch2020kinetic}
to model observed systematic 
asymmetries of particle density and temperature
across kinetic scale solar wind current sheets.
In the previous approach, both ion and
electron distribution functions were modified
from a known force-free case 
\citep[e.g.][]{harrison2009one,Neukirch-2009}
to maintain exact charge neutrality leading to 
a vanishing electric potential. Here we have
investigated the consequences of modifying
only the electron distribution function.
This leads to a system of equations in which
a nonlinear quasineutrality condition 
coupling the electric potential and the
vector potential has to be solved alongside
Amp\`{e}re's law. We showed that a ``crude"
approximation method in which only the electric
potential is expanded in terms of a small
parameter, while the 
force-free magnetic field is kept 
unchanged 
\citep[][]{neukirch2022kinetic},
is inconsistent. We have presented results
of both a consistent
expansion method which includes the magnetic field
(up to linear terms)
and of solving the full nonlinear problem.
These results 
showed 
that while the changes in the magnetic field are
relatively small, the quasineutral electric potential 
has an unexpected spatial substructure inside the
current sheet, which in turn leads to a similar
spatial substructure of the particle density.

While this is an interesting theoretical result,
it is 
different from the 
monotonically varying
particle
density (and temperature) profiles 
found
by
\citet{neukirch2020kinetic}.
One immediate question that arises is whether
there are any observational examples at all
that display a spatial substructure
of a similar nature to our findings
is this paper. 
We do not claim that the following constitutes
a proper comparison of our theory
with the observations. 
 However, there seem to be examples of current sheet observations in the 
 solar wind  that 
 display substructures not too dissimilar to those found in this study.

\begin{figure} 
\centerline{\includegraphics[width=\textwidth,clip=]{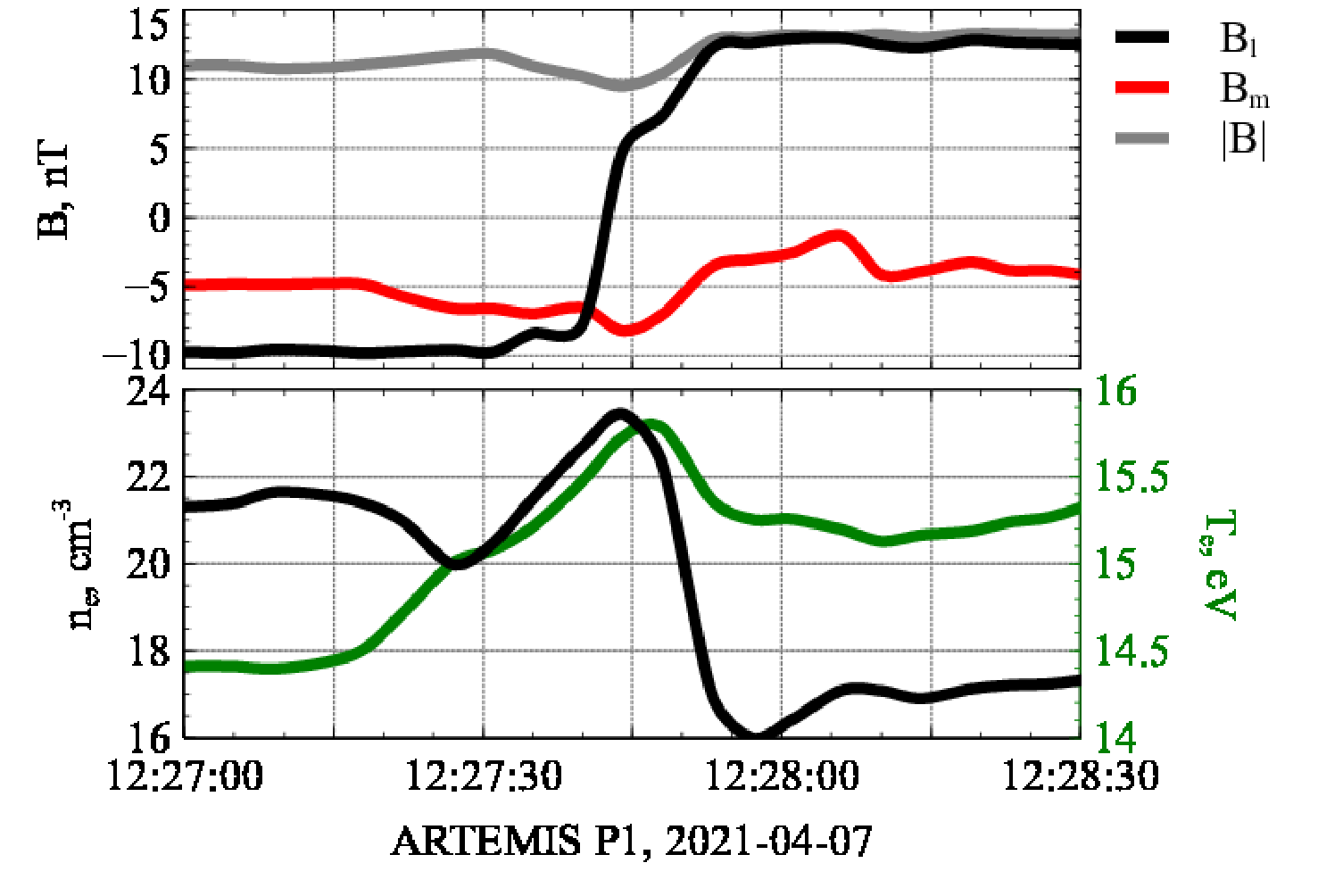}}
\vspace{-0.67\textwidth}   
\centerline{\Large \bf     
\hspace{0 \textwidth}  \color{black}{(a)}
   \hfill}
\vspace{0.25\textwidth}   
\centerline{\Large \bf     
\hspace{0 \textwidth}  \color{black}{(b)}
   \hfill}
\vspace{0.32\textwidth}

        \caption{An example of a current sheet observed by the ARTEMIS mission near 1 AU \citep[][]{Angelopoulos:2011:ssr} showing (a) the magnetic field components in the local coordinate system, and (b) the plasma (electron) density and electron temperature profiles. We remark that the theoretical
        result presented before show spatial variations of the electron density in the opposite direction to the
        observation shown here but that is merely a consequence of the symmetry chosen for the theoretical calculations which could be reversed without problem.}
\label{F-ex_obs}
\end{figure}

Figure 
\ref{F-ex_obs} shows one such example observed by the ARTEMIS mission near 1 AU 
\citep[][]{Angelopoulos:2011:ssr}. 
The top panel presents the magnetic field measurements from the fluxgate magnetometer 
\citep[][]{Auster:etal2008:ssr}. A local coordinate system is used, with $B_l\approx B_x$ and $B_m \approx B_y$ 
\citep[for details on current sheet selection and magnetic field processing, see][]{Artemyev-2019b}. The magnetic field configuration exhibits a $B_l$ reversal and a peak in $|B_m|$, indicative of a nearly constant magnetic field magnitude ($B \approx const$) across a force-free current sheet. Since solar wind current sheets are convected by plasma flows, the time series can be interpreted in terms of spatial scales 
\citep[see, e.g.][]{Artemyev-2019b,neukirch2020kinetic}. The bottom panel shows the plasma (electron) density and electron temperature profiles. In this particular current sheet, the ion temperature is approximately half that of the electron temperature. Superimposed on the general trends—an increase in density and a decrease in temperature—are non-monotonic substructures near the current sheet center (at the $B_l$ reversal), which resemble those predicted by the theoretical model. Although a more detailed analysis of plasma kinetics, including the velocity distribution functions and their evolution across the current sheet, is necessary for a quantitative comparison, the example in Figure 
\ref{F-ex_obs} illustrates the potential for such observational–theoretical correspondence.

We have not carried out a full systematic comparison here, and our results are not the only possible explanation of the observations shown. In particular it will be important to assess the full details of pressure balance in future work (beyond the scope of the work presented here). However, it is important to note the observations of current sheets in the solar wind that have a similar substructure to our theoretical findings.

\begin{acks}
S.B. acknowledges support from the United Kingdom Research and Innovation (UKRI) Science and Technology Facilities
Council (STFC) Doctoral Training Partnership Grant
ST/X508779/1. T.N. acknowledges support by STFC Consolidated
Grants ST/S000402/1 and ST/W001195/1.
A.A. acknowledges support from NASA 80NSSC23K0658, 80NSSC22K0752.
 I.V. thanks RSF project No. 24-12-00457.
 O.A. would like to acknowledge support from the University of Birmingham, and also from the United Kingdom Research and Innovation (UKRI) Natural Environment Research Council (NERC) Independent Research Fellowship NE/V013963/1 and NE/V013963/2.
\end{acks}



\appendix

\section{Basic Current Sheet Models: Further Details}
\label{S-appendix-CS}
 
For reference, we briefly summarise in this appendix further details of
one-dimensional collisionless current sheet equilibrium theory. 
For the coordinate system chosen 
in this paper a crucial quantity is the
$zz$-component of the plasma pressure tensor, which is defined as
\begin{equation}
    P_{zz}(\Phi, A_x, A_y)= \sum_s m_s \int v_z^2 F_s(H_s, p_{x,s}, p_{y,s})\, \mathrm{d}^3 v.
    \label{eq:Pzz-def}
\end{equation}

It can be shown \citep[see e.g.][]{Mynick:etal79,harrison2009some} that
\begin{eqnarray}
    \sigma &=& -\frac{\partial P_{zz}}{\partial \Phi} ,  \label{eq:cd-dPdPhi} \\
    j_x    &=& \frac{\partial P_{zz}}{\partial A_x} ,  \label{eq:jx-dPdAx} \\
    j_y    &=& \frac{\partial P_{zz}}{\partial A_y} ,  \label{eq:jy-dPdAy}
\end{eqnarray}
which makes it very convenient to use 
$P_{zz}$ to determine the charge density, $\sigma$,
and the components of the current density, $j_x$ and $j_y$.
For the assumptions made in this paper, a self-consistent
one-dimensional quasineutral
collisionless current sheet automatically satisfies the force balance equation,
which is the first order velocity moment of the
stationary Vlasov equation \citep[see e.g.][]{Mynick:etal79}. The
only non-trivial component of the force balance equation is the $z$-component,
given by
\begin{equation}
    \frac{\mathrm{d}}{\mathrm{d} z}\left( \frac{B_x^2 +B_y^2}{2\mu_0} + P_{zz}\right) =0.
    \label{eq:forcebalance-general}
\end{equation}

For a one-dimensional force-free current sheet the force balance equation splits
into the two conditions \citep[see e.g.][]{harrison2009some}
\begin{eqnarray}
    \frac{\mathrm{d}}{\mathrm{d} z}\left( \frac{B_x^2 +B_y^2}{2\mu_0} \right) &=& 0,
    \label{eq:ff-B2} \\
    \frac{\mathrm{d} P_{zz}}{\mathrm{d} z} &=& 0,
    \label{eq:ff-DPdz}
\end{eqnarray}
implying that both $B^2$ and $P_{zz}$ are constant for a one-dimensional force-free 
current sheet.

The contributions to the  $zz$-component of the pressure tensor resulting from the
electron distribution function given by Equation \ref{DFe} and 
the ion distribution function given by Equation \ref{DFi}, 
respectively, are 
\begin{eqnarray}
    P_{e,ff} &=& \frac{n_0}{\beta_e} e^{e\beta_{e}\Phi}
                N(A_x,A_y),
                           \label{eq:Pzz_eff} \\
    P_{i,ff} &=& \frac{n_{0,i}}{\beta_i} e^{-e\beta_{i}\Phi}     ,         \label{eq:Pzz_iff}       
\end{eqnarray}
where $N(A_x,A_y)$ is defined in 
Equation \ref{eq-defN}.
The derivatives with
respect to $A_x$ and $A_y$ of $P_{e,ff}$ ($P_{i,ff}$
does not depend on the vector potential) result in the 
first terms on the right hand sides of Amp\`{e}re's law for 
Equations \ref{Ampx}
and \ref{Ampy} for $\epsilon =0$. 
Setting the derivative with respect to $\Phi$ equal to zero results in the quasineutrality condition for the normal force-free case
\begin{equation}
    n_{0,i} e^{-e\beta_{i}\Phi} - n_0 e^{e\beta_{e}\Phi}
    N(A_x,A_y) = 0
    \label{eq:QN-ff}
\end{equation}
Using standard 
identities it is 
straightforward to show that
$N(A_{x,ff}, A_{y,ff}) = b+1/2$ (
Equation \ref{rel1})
\citep[e.g.][]{harrison2009one,Neukirch-2009,Neukirch-2017}
and that therefore 
Equation \ref{eq:QN-ff} is solved by 
$\Phi = 0$, with $n_{0,i} = n_0(b+ 1/2)$ (
Equation \ref{rel2}).

\begin{figure}
\centerline{\hspace*{0.015\textwidth}
         \includegraphics[width=0.515\textwidth,clip=]{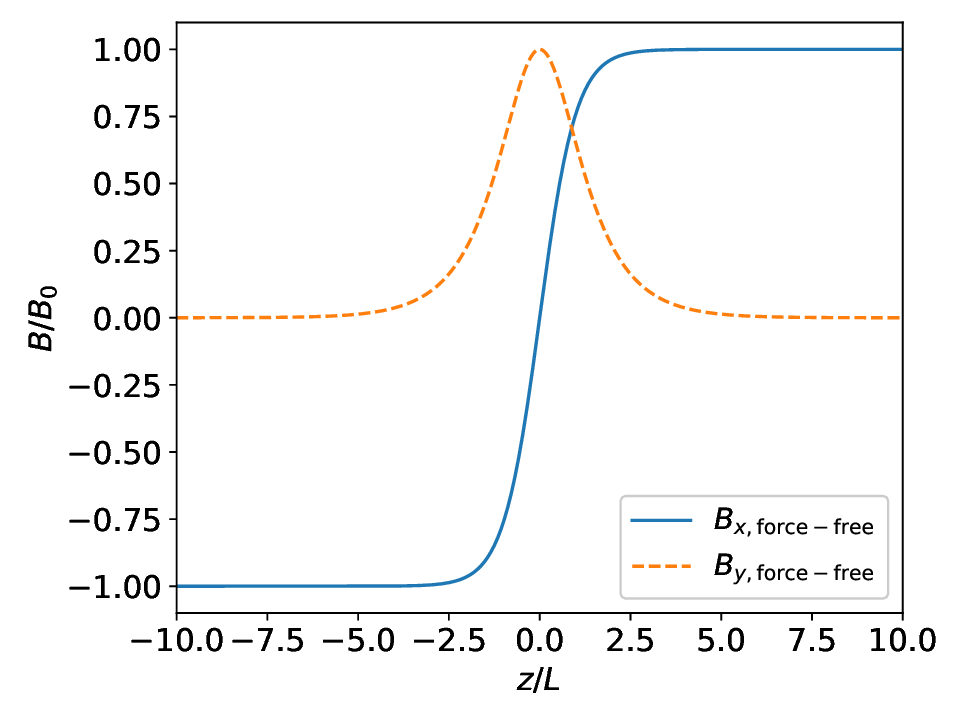}
         \hspace*{-0.03\textwidth}
         \includegraphics[width=0.515\textwidth,clip=]{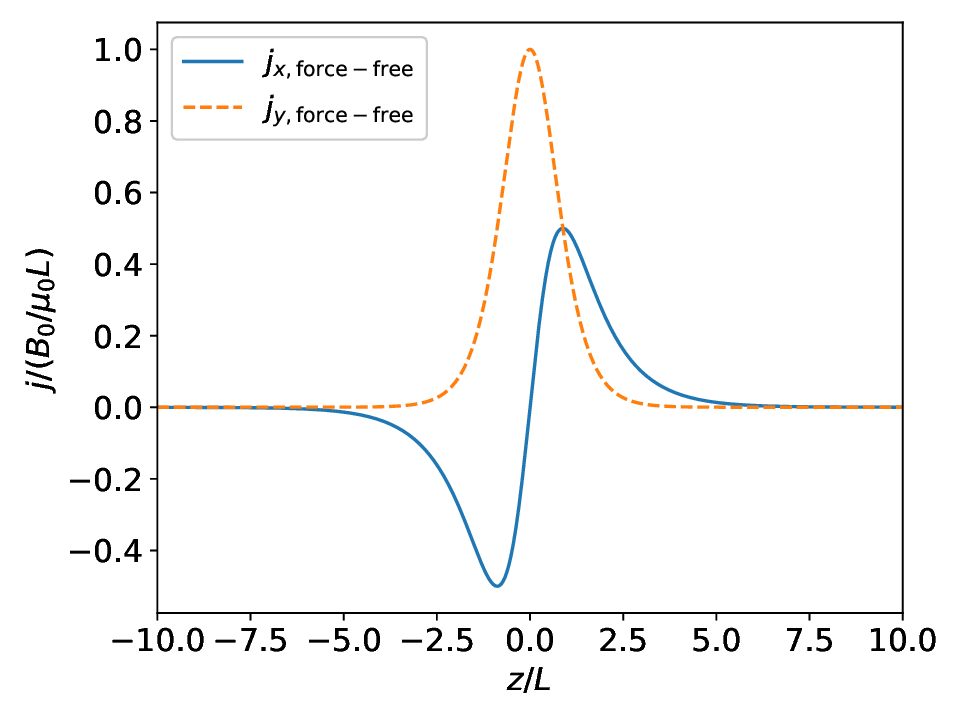}
        }
\vspace{-0.351\textwidth}   
\centerline{\Large \bf     
\hspace{0.375 \textwidth}  \color{black}{(a)}
\hspace{0.415\textwidth}  \color{black}{(b)}
   \hfill}
\vspace{0.31\textwidth}    
\centerline{\hspace*{0.015\textwidth}
         \includegraphics[width=0.515\textwidth,clip=]{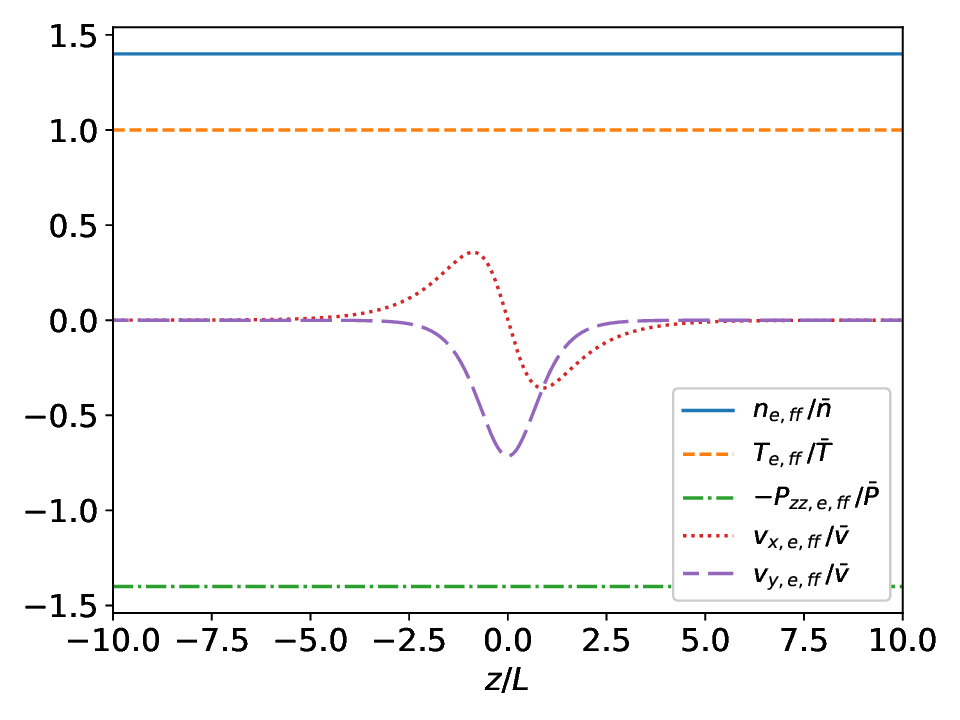}
         \hspace*{-0.03\textwidth}
         \includegraphics[width=0.515\textwidth,clip=]{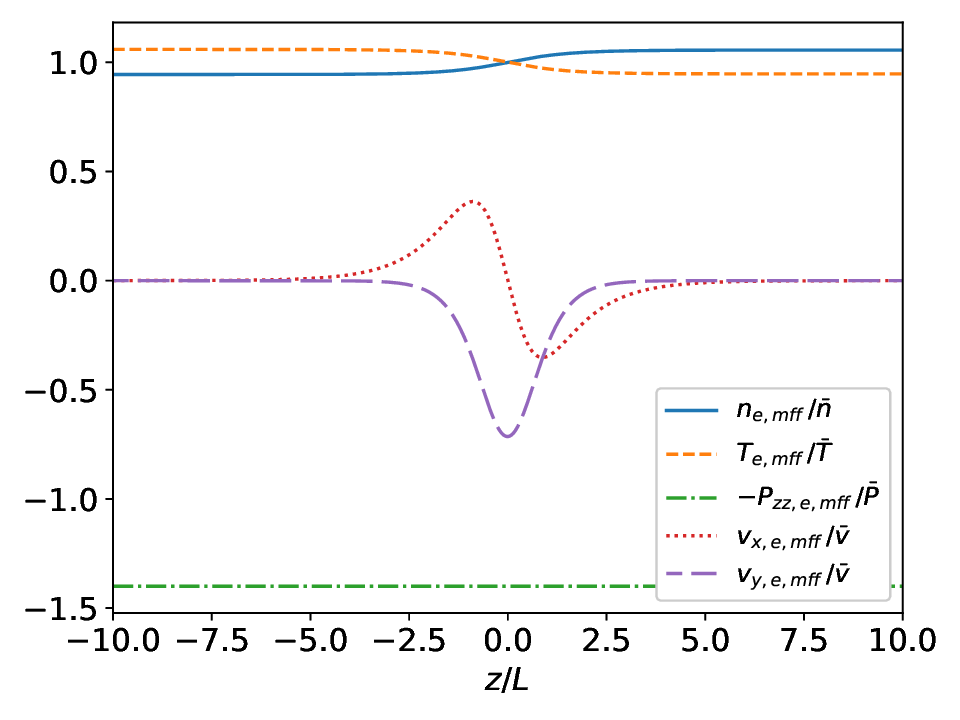}
        }
\vspace{-0.351\textwidth}   
\centerline{\Large \bf     
\hspace{0.375 \textwidth} \color{black}{(c)}
\hspace{0.415\textwidth}  \color{black}{(d)}
   \hfill}
\vspace{0.31\textwidth}    
              
\caption{Profiles of the force-free (a) magnetic field and (b) current density, and the electron density, temperature, pressure, and drift velocity for the (c) force-free case (labelled `ff' in the legend) and (d) modified force-free case (labelled `mff' in the legend). Note that $-P_{zz,e}$ has been plotted since $P_{zz,e}$ has the same profile as $n_{e}$ in the force-free case. We also note that the normalising quantities in panels (c) and (d) are as follows: $\bar{n}=n_{0}(b+1/2)$, $\bar{P}=B_{0}^2/\mu_{0}$, $\bar{T}=B_{0}^2/(k_{B}\mu_{0}n_{0}(b+1/2))$, and $\bar{v}=\left|u_{0}\right|/(b+1/2)$.}
\label{F-forcefree}
\end{figure}

The current density resulting from the 
force-free magnetic field in 
Equation \ref{magfield} is
\begin{equation}
 \mu_0   \mathbf{j}_{ff}(z) = \frac{B_0}{L \cosh^2(z/L)} [\sinh(z/L), 1, 0].
    \label{eq:ff-currentdensity}
\end{equation}
Plots of the force-free magnetic field and current density components as well as the electron density, $zz$-component
of the electron pressure tensor, electron temperature and
electron drift velocity (for both force-free and the modified force-free case) are
shown in Figure \ref{F-forcefree}.
For comparison, we have also included plots
of the magnetic field and current density in Figures \ref{F-linear} (panels (a) and (b), respectively) and
\ref{F-nonlinear} (panels (a) and (b)).
Again, one can easily 
verify analytically that $j_{x,ff}(z) = en_{0}u_{0}\frac{1}{2}\sin(e\beta_{e}u_{0}A_{x,ff})$
and $j_{y,ff}(z) = en_{0}u_{0}e^{-e\beta_{e}u_{0}A_{y,ff}}$
if the relations given by Equations \ref{rel3} - \ref{rel6} hold 
\citep[e.g.][]{harrison2009one,Neukirch-2009,Neukirch-2017}.

The additional distribution function given by Equation \ref{DFaddterm}
results in additional terms for the $zz$-component of
the pressure tensor for particle species $s$ of the form
\begin{equation}
    P_{\Delta,s} = - \delta n_s q_s^2 \beta_s u_0 A_x \Phi e^{-\kappa_s \beta_s q_s \Phi}.
    \label{eq:P_delta-s}
\end{equation}
The additional terms 
for
the charge density and hence
the quasineutrality condition can be calculated using
Equation \ref{eq:cd-dPdPhi}. Using the relation given by Equation \ref{rel3}
and $\delta n_e = - 
\delta n_i \beta_i/\beta_e=
\epsilon n_0$, one obtains
\begin{eqnarray}
    \Delta \sigma_e &= &- e \epsilon n_0 \frac{2 A_x}{B_0 L}
    (1 + \kappa_e \beta_e e \Phi) e^{\kappa_e \beta_e e \Phi},
    \label{eq:delta-sigma_e} \\
    \Delta \sigma_i &=&  e \epsilon n_0 \frac{2 A_x}{B_0 L}
    \left(1 - \kappa_i \beta_i e \Phi\right) 
    e^{-\kappa_i \beta_i e \Phi}.
    \label{eq:delta-sigma_i}
\end{eqnarray}
It is obvious that $\Phi =0 $ is still a solution of the
quasineutrality equation even if the term
$\Delta \sigma_e + \Delta \sigma_i$ is added. 

Because the additional pressure tensor term only depends
on $A_x$ according to 
Equation \ref{eq:jx-dPdAx} it will
only contribute to $j_x$. The partial derivative
with respect to $A_x$ of $P_{\Delta,s}$ is given
by
\begin{equation}
    \frac{\partial P_{\Delta,s}}{\partial A_x} 
    = - q_s \delta n_s u_0 (\beta_s q_s \Phi) 
    e^{-\kappa_s \beta_s q_s \Phi },
    \label{eq:jx-Delta}
\end{equation}
which vanishes for $\Phi=0$. Hence for the case
in which both the electron and 
the ion distribution functions are modified the 
current density remains unchanged and
hence the force-free magnetic field stated in
Equation \ref{magfield} remains a self-consistent solution
of the modified problem.

In all of the cases discussed 
before (purely force-free and modified
force-free) or in this paper the electric current density
is carried by the electrons only, which also
implies that only the electrons have 
a non-vanishing average velocity, which is 
identical to the current density divided by the
electron charge density. Like all quantities
the average velocity only depends on $z$, and since only its $x$- and $y$-components are non-zero it does not contribute
in any way to the force balance of the system.

\section{Reformulated Nonlinear Differential Equations} 
\label{S-appendix}

Using the differences $\Delta \bar{A}_x$ and $\Delta \bar{A}_y$  between the vector potential components and their force-free counterparts defined by $\bar{A}_x = \bar{A}_{x,ff} + \Delta \bar{A}_x$, 
$\bar{A}_y = \bar{A}_{y,ff} + \Delta \bar{A}_y$, Equations \ref{normAmpx}, \ref{normAmpy} and \ref{dPhidz} are reformulated as

\begin{eqnarray}
    \frac{d^{2}\Delta\bar{A}_{x}}{d\bar{z}^{2}} &=& \frac{\sinh(\bar{z})}{\cosh^{2}(\bar{z})}\left(-e^{\bar{\Phi}}\cos(2\Delta\bar{A}_{x}) + 1\right) - \frac{2 - \cosh^{2}(\bar{z})}{2\cosh^{2}(\bar{z})}e^{\bar{\Phi}}\sin(2\Delta\bar{A}_{x})\nonumber\\&&\hspace{20em} - \epsilon\bar{\Phi}e^{\kappa_{e}\bar{\Phi}},\label{-jx_reform}\\
    \frac{d^{2}\Delta\bar{A}_{y}}{d\bar{z}^{2}} &=& \frac{1}{\cosh^{2}(\bar{z})}\left(-e^{\bar{\Phi}}e^{2\Delta\bar{A}_{y}} + 1\right),\label{-jy_reform}\\
    \frac{d\bar{\Phi}}{d\bar{z}} &=& \biggl\{\biggl(\frac{2\sinh(\bar{z})}{\cosh^{2}(\bar{z})}e^{\bar{\Phi}}\cos(2\Delta\bar{A}_{x}) + \frac{2-\cosh^{2}(\bar{z})}{\cosh^{2}(\bar{z})}e^{\bar{\Phi}}\sin(2\Delta\bar{A}_{x})\nonumber \\ &&\hspace{1.5em} +2\epsilon(1+\kappa_{e}\bar{\Phi})e^{\kappa_{e}\bar{\Phi}}\biggr)\cdot\left(\frac{1}{\cosh(\bar{z})}+\frac{d\Delta\bar{A}_{x}}{d\bar{z}}\right)\nonumber\\&&\hspace{2.5em} + \left(\frac{2}{\cosh^{2}(\bar{z})}e^{\bar{\Phi}}e^{2\Delta\bar{A}_{y}}\right) \cdot\left(-\frac{\sinh(\bar{z})}{\cosh(\bar{z})}+\frac{d\Delta\bar{A}_{y}}{d\bar{z}}\right)\biggr\}\nonumber \\ &&\hspace{3.5em}\cdot\biggl\{-\frac{\beta_{i}}{\beta_{e}}\left(b+\frac{1}{2}\right)e^{-\frac{\beta_{i}}{\beta_{e}}\bar{\Phi}}  -e^{\bar{\Phi}}\left( \frac{\sinh(\bar{z})}{\cosh^{2}(\bar{z})}\sin(2\Delta\bar{A}_{x})\right.\nonumber \\&&\hspace{4.5em}\left. - \frac{2-\cosh^{2}(\bar{z})}{2\cosh^{2}(\bar{z})}\cos(2\Delta\bar{A}_{x}) + \frac{1}{\cosh^{2}(\bar{z})}e^{2\Delta\bar{A}_{y}} +b\right)\nonumber \\&&\hspace{5.5em} -2\epsilon\kappa_{e}(2+\kappa_{e}\bar{\Phi})e^{\kappa_{e}\bar{\Phi}}(\bar{A}_{x,ff}+\Delta\bar{A}_{x})\biggr\}^{-1}.\label{dPhi_reform}
\end{eqnarray}




\begin{authorcontribution}
S.B. carried out the numerical calculations with
assistance by T.N. All authors contributed to the
formulation of the problem, the interpretation of the results and the writing of the paper.
\end{authorcontribution}

\begin{fundinginformation}
This work was supported by the UK's Science and Technology
Facilities Council (STFC) via Doctoral Training Partnership Grant ST/X508779/1, Consolidated Grants ST/S000402/1 and ST/W001195/1, the UK's Natural Environment Research Council (NERC) Independent Research Fellowship
Grants NE/V013963/1 and NE/V013963/2, RSF project No. 24-12-00457, and NASA 80NSSC23K0658 and 80NSSC22K0752.
\end{fundinginformation}

\begin{dataavailability}
ARETMIS data are available at \url{http://themis.ssl.berkeley.edu}. Data was retrieved and analyzed using SPEDAS, see \citet[][]{Angelopoulos:etal2019:ssr}.
\end{dataavailability}



\begin{ethics}
\begin{conflict}
The authors declare that they have no conflicts of interest.
\end{conflict}
\end{ethics}
  
\bibliographystyle{spr-mp-sola}
\bibliography{SB-et-al-refs}

\IfFileExists{\jobname.bbl}{} {\typeout{}
\typeout{****************************************************}
\typeout{****************************************************}
\typeout{** Please run "bibtex \jobname" to obtain} \typeout{**
the bibliography and then re-run LaTeX} \typeout{** twice to fix
the references !}
\typeout{****************************************************}
\typeout{****************************************************}
\typeout{}}

\end{document}